\documentclass[10pt,conference,twoside]{IEEEtran}

\usepackage{cite}
\usepackage[T1]{fontenc}
\usepackage{graphicx}
\usepackage{amssymb}
\usepackage{amsmath}
\usepackage{amsthm}

\usepackage{microtype}
\usepackage{hyperref}
\usepackage{url}
\usepackage{balance}
\usepackage{xcolor}
\usepackage{subfigure}

\usepackage{amssymb}
\usepackage{amsfonts}
\usepackage{mathrsfs}
\usepackage{xspace}
\usepackage{bm}
\usepackage{upgreek}

\newcommand{\safemath}[2]{\newcommand{#1}{\ensuremath{#2}\xspace}}

\safemath{\bma}{\mathbf{a}}
\safemath{\bmb}{\mathbf{b}}
\safemath{\bmc}{\mathbf{c}}
\safemath{\bmd}{\mathbf{d}}
\safemath{\bme}{\mathbf{e}}
\safemath{\bmf}{\mathbf{f}}
\safemath{\bmg}{\mathbf{g}}
\safemath{\bmh}{\mathbf{h}}
\safemath{\bmi}{\mathbf{i}}
\safemath{\bmj}{\mathbf{j}}
\safemath{\bmk}{\mathbf{k}}
\safemath{\bml}{\mathbf{l}}
\safemath{\bmm}{\mathbf{m}}
\safemath{\bmn}{\mathbf{n}}
\safemath{\bmo}{\mathbf{o}}
\safemath{\bmp}{\mathbf{p}}
\safemath{\bmq}{\mathbf{q}}
\safemath{\bmr}{\mathbf{r}}
\safemath{\bms}{\mathbf{s}}
\safemath{\bmt}{\mathbf{t}}
\safemath{\bmu}{\mathbf{u}}
\safemath{\bmv}{\mathbf{v}}
\safemath{\bmw}{\mathbf{w}}
\safemath{\bmx}{\mathbf{x}}
\safemath{\bmy}{\mathbf{y}}
\safemath{\bmz}{\mathbf{z}}
\safemath{\bmzero}{\mathbf{0}}
\safemath{\bmone}{\mathbf{1}}

\bmdefine{\biad}{a}
\bmdefine{\bibd}{b}
\bmdefine{\bicd}{c}
\bmdefine{\bidd}{d}
\bmdefine{\bied}{e}
\bmdefine{\bifd}{f}
\bmdefine{\bigd}{g}
\bmdefine{\bihd}{h}
\bmdefine{\biid}{i}
\bmdefine{\bijd}{j}
\bmdefine{\bikd}{k}
\bmdefine{\bild}{l}
\bmdefine{\bimd}{m}
\bmdefine{\bind}{n}
\bmdefine{\biod}{o}
\bmdefine{\bipd}{p}
\bmdefine{\biqd}{q}
\bmdefine{\bird}{r}
\bmdefine{\bisd}{s}
\bmdefine{\bitd}{t}
\bmdefine{\biud}{u}
\bmdefine{\bivd}{v}
\bmdefine{\biwd}{w}
\bmdefine{\bixd}{x}
\bmdefine{\biyd}{y}
\bmdefine{\bizd}{z}

\bmdefine{\bixid}{\xi}
\bmdefine{\bilambdad}{\lambda}
\bmdefine{\bimud}{\mu}
\bmdefine{\bithetad}{\theta}
\bmdefine{\biphid}{\phi}
\bmdefine{\bideltad}{\delta}

\safemath{\bmia}{\biad}
\safemath{\bmib}{\bibd}
\safemath{\bmic}{\bicd}
\safemath{\bmid}{\bidd}
\safemath{\bmie}{\bied}
\safemath{\bmif}{\bifd}
\safemath{\bmig}{\bigd}
\safemath{\bmih}{\bihd}
\safemath{\bmii}{\biid}
\safemath{\bmij}{\bijd}
\safemath{\bmik}{\bikd}
\safemath{\bmil}{\bild}
\safemath{\bmim}{\bimd}
\safemath{\bmin}{\bind}
\safemath{\bmio}{\biod}
\safemath{\bmip}{\bipd}
\safemath{\bmiq}{\biqd}
\safemath{\bmir}{\bird}
\safemath{\bmis}{\bisd}
\safemath{\bmit}{\bitd}
\safemath{\bmiu}{\biud}
\safemath{\bmiv}{\bivd}
\safemath{\bmiw}{\biwd}
\safemath{\bmix}{\bixd}
\safemath{\bmiy}{\biyd}
\safemath{\bmiz}{\bizd}

\safemath{\bmxi}{\bixid}
\safemath{\bmlambda}{\bilambdad}
\safemath{\bmmu}{\bimud}
\safemath{\bmtheta}{\bithetad}
\safemath{\bmphi}{\biphid}
\safemath{\bmdelta}{\bideltad}

\safemath{\bA}{\mathbf{A}}
\safemath{\bB}{\mathbf{B}}
\safemath{\bC}{\mathbf{C}}
\safemath{\bD}{\mathbf{D}}
\safemath{\bE}{\mathbf{E}}
\safemath{\bF}{\mathbf{F}}
\safemath{\bG}{\mathbf{G}}
\safemath{\bH}{\mathbf{H}}
\safemath{\bI}{\mathbf{I}}
\safemath{\bJ}{\mathbf{J}}
\safemath{\bK}{\mathbf{K}}
\safemath{\bL}{\mathbf{L}}
\safemath{\bM}{\mathbf{M}}
\safemath{\bN}{\mathbf{N}}
\safemath{\bO}{\mathbf{O}}
\safemath{\bP}{\mathbf{P}}
\safemath{\bQ}{\mathbf{Q}}
\safemath{\bR}{\mathbf{R}}
\safemath{\bS}{\mathbf{S}}
\safemath{\bT}{\mathbf{T}}
\safemath{\bU}{\mathbf{U}}
\safemath{\bV}{\mathbf{V}}
\safemath{\bW}{\mathbf{W}}
\safemath{\bX}{\mathbf{X}}
\safemath{\bY}{\mathbf{Y}}
\safemath{\bZ}{\mathbf{Z}}

\safemath{\bZero}{\mathbf{0}}
\safemath{\bOne}{\mathbf{1}}
\safemath{\bDelta}{\mathbf{\Delta}}
\safemath{\bLambda}{\mathbf{\UpLambda}}
\safemath{\bPhi}{\mathbf{\Upphi}}
\safemath{\bSigma}{\mathbf{\Upsigma}}
\safemath{\bOmega}{\mathbf{\Upomega}}
\safemath{\bTheta}{\mathbf{\Uptheta}}

\bmdefine{\biAd}{A}
\bmdefine{\biBd}{B}
\bmdefine{\biCd}{C}
\bmdefine{\biDd}{D}
\bmdefine{\biEd}{E}
\bmdefine{\biFd}{F}
\bmdefine{\biGd}{G}
\bmdefine{\biHd}{H}
\bmdefine{\biId}{I}
\bmdefine{\biJd}{J}
\bmdefine{\biKd}{K}
\bmdefine{\biLd}{L}
\bmdefine{\biMd}{M}
\bmdefine{\biOd}{N}
\bmdefine{\biPd}{O}
\bmdefine{\biQd}{P}
\bmdefine{\biRd}{R}
\bmdefine{\biSd}{S}
\bmdefine{\biTd}{T}
\bmdefine{\biUd}{U}
\bmdefine{\biVd}{V}
\bmdefine{\biWd}{W}
\bmdefine{\biXd}{X}
\bmdefine{\biYd}{Y}
\bmdefine{\biZd}{Z}

\bmdefine{\biDelta}{\Delta}
\bmdefine{\biLambda}{\Lambda}
\bmdefine{\biPhi}{\Phi}
\bmdefine{\biSigma}{\Sigma}
\bmdefine{\biOmega}{\Omega}
\bmdefine{\biTheta}{\Theta}

\safemath{\bimA}{\biAd}
\safemath{\bimB}{\biBd}
\safemath{\bimC}{\biCd}
\safemath{\bimD}{\biDd}
\safemath{\bimE}{\biEd}
\safemath{\bimF}{\biFd}
\safemath{\bimG}{\biGd}
\safemath{\bimH}{\biHd}
\safemath{\bimI}{\biId}
\safemath{\bimJ}{\biJd}
\safemath{\bimK}{\biKd}
\safemath{\bimL}{\biLd}
\safemath{\bimM}{\biMd}
\safemath{\bimN}{\biNd}
\safemath{\bimO}{\biOd}
\safemath{\bimP}{\biPd}
\safemath{\bimQ}{\biQd}
\safemath{\bimR}{\biRd}
\safemath{\bimS}{\biSd}
\safemath{\bimT}{\biTd}
\safemath{\bimU}{\biUd}
\safemath{\bimV}{\biVd}
\safemath{\bimW}{\biWd}
\safemath{\bimX}{\biXd}
\safemath{\bimY}{\biYd}
\safemath{\bimZ}{\biZd}

\safemath{\bimDelta}{\biDelta}
\safemath{\bimLambda}{\biLambda}
\safemath{\bimPhi}{\biPhi}
\safemath{\bimSigma}{\biSigma}
\safemath{\bimOmega}{\biOmega}
\safemath{\bimTheta}{\biTheta}

\safemath{\setA}{\mathcal{A}}
\safemath{\setB}{\mathcal{B}}
\safemath{\setC}{\mathcal{C}}
\safemath{\setD}{\mathcal{D}}
\safemath{\setE}{\mathcal{E}}
\safemath{\setF}{\mathcal{F}}
\safemath{\setG}{\mathcal{G}}
\safemath{\setH}{\mathcal{H}}
\safemath{\setI}{\mathcal{I}}
\safemath{\setJ}{\mathcal{J}}
\safemath{\setK}{\mathcal{K}}
\safemath{\setL}{\mathcal{L}}
\safemath{\setM}{\mathcal{M}}
\safemath{\setN}{\mathcal{N}}
\safemath{\setO}{\mathcal{O}}
\safemath{\setP}{\mathcal{P}}
\safemath{\setQ}{\mathcal{Q}}
\safemath{\setR}{\mathcal{R}}
\safemath{\setS}{\mathcal{S}}
\safemath{\setT}{\mathcal{T}}
\safemath{\setU}{\mathcal{U}}
\safemath{\setV}{\mathcal{V}}
\safemath{\setW}{\mathcal{W}}
\safemath{\setX}{\mathcal{X}}
\safemath{\setY}{\mathcal{Y}}
\safemath{\setZ}{\mathcal{Z}}
\safemath{\emptySet}{\varnothing}

\safemath{\colA}{\mathscr{A}}
\safemath{\colB}{\mathscr{B}}
\safemath{\colC}{\mathscr{C}}
\safemath{\colD}{\mathscr{D}}
\safemath{\colE}{\mathscr{E}}
\safemath{\colF}{\mathscr{F}}
\safemath{\colG}{\mathscr{G}}
\safemath{\colH}{\mathscr{H}}
\safemath{\colI}{\mathscr{I}}
\safemath{\colJ}{\mathscr{J}}
\safemath{\colK}{\mathscr{K}}
\safemath{\colL}{\mathscr{L}}
\safemath{\colM}{\mathscr{M}}
\safemath{\colN}{\mathscr{N}}
\safemath{\colO}{\mathscr{O}}
\safemath{\colP}{\mathscr{P}}
\safemath{\colQ}{\mathscr{Q}}
\safemath{\colR}{\mathscr{R}}
\safemath{\colS}{\mathscr{S}}
\safemath{\colT}{\mathscr{T}}
\safemath{\colU}{\mathscr{U}}
\safemath{\colV}{\mathscr{V}}
\safemath{\colW}{\mathscr{W}}
\safemath{\colX}{\mathscr{X}}
\safemath{\colY}{\mathscr{Y}}
\safemath{\colZ}{\mathscr{Z}}

\safemath{\opA}{\mathbb{A}}
\safemath{\opB}{\mathbb{B}}
\safemath{\opC}{\mathbb{C}}
\safemath{\opD}{\mathbb{D}}
\safemath{\opE}{\mathbb{E}}
\safemath{\opF}{\mathbb{F}}
\safemath{\opG}{\mathbb{G}}
\safemath{\opH}{\mathbb{H}}
\safemath{\opI}{\mathbb{I}}
\safemath{\opJ}{\mathbb{J}}
\safemath{\opK}{\mathbb{K}}
\safemath{\opL}{\mathbb{L}}
\safemath{\opM}{\mathbb{M}}
\safemath{\opN}{\mathbb{N}}
\safemath{\opO}{\mathbb{O}}
\safemath{\opP}{\mathbb{P}}
\safemath{\opQ}{\mathbb{Q}}
\safemath{\opR}{\mathbb{R}}
\safemath{\opS}{\mathbb{S}}
\safemath{\opT}{\mathbb{T}}
\safemath{\opU}{\mathbb{U}}
\safemath{\opV}{\mathbb{V}}
\safemath{\opW}{\mathbb{W}}
\safemath{\opX}{\mathbb{X}}
\safemath{\opY}{\mathbb{Y}}
\safemath{\opZ}{\mathbb{Z}}
\safemath{\opZero}{\mathbb{O}}
\safemath{\identityop}{\opI}

\safemath{\veca}{\bma}
\safemath{\vecb}{\bmb}
\safemath{\vecc}{\bmc}
\safemath{\vecd}{\bmd}
\safemath{\vece}{\bme}
\safemath{\vecf}{\bmf}
\safemath{\vecg}{\bmg}
\safemath{\vech}{\bmh}
\safemath{\veci}{\bmi}
\safemath{\vecj}{\bmj}
\safemath{\veck}{\bmk}
\safemath{\vecl}{\bml}
\safemath{\vecm}{\bmm}
\safemath{\vecn}{\bmn}
\safemath{\veco}{\bmo}
\safemath{\vecp}{\bmp}
\safemath{\vecq}{\bmq}
\safemath{\vecr}{\bmr}
\safemath{\vecs}{\bms}
\safemath{\vect}{\bmt}
\safemath{\vecu}{\bmu}
\safemath{\vecv}{\bmv}
\safemath{\vecw}{\bmw}
\safemath{\vecx}{\bmx}
\safemath{\vecy}{\bmy}
\safemath{\vecz}{\bmz}

\safemath{\veczero}{\bmzero}
\safemath{\vecone}{\bmone}
\safemath{\vecxi}{\bmxi}
\safemath{\veclambda}{\bmlambda}
\safemath{\vecmu}{\bmmu}
\safemath{\vectheta}{\bmtheta}
\safemath{\vecphi}{\bmphi}
\safemath{\vecdelta}{\bmdelta}

\safemath{\matA}{\bA}
\safemath{\matB}{\bB}
\safemath{\matC}{\bC}
\safemath{\matD}{\bD}
\safemath{\matE}{\bE}
\safemath{\matF}{\bF}
\safemath{\matG}{\bG}
\safemath{\matH}{\bH}
\safemath{\matI}{\bI}
\safemath{\matJ}{\bJ}
\safemath{\matK}{\bK}
\safemath{\matL}{\bL}
\safemath{\matM}{\bM}
\safemath{\matN}{\bN}
\safemath{\matO}{\bO}
\safemath{\matP}{\bP}
\safemath{\matQ}{\bQ}
\safemath{\matR}{\bR}
\safemath{\matS}{\bS}
\safemath{\matT}{\bT}
\safemath{\matU}{\bU}
\safemath{\matV}{\bV}
\safemath{\matW}{\bW}
\safemath{\matX}{\bX}
\safemath{\matY}{\bY}
\safemath{\matZ}{\bZ}
\safemath{\matzero}{\bmzero}

\safemath{\matDelta}{\bDelta}
\safemath{\matLambda}{\bLambda}
\safemath{\matPhi}{\bPhi}
\safemath{\matSigma}{\bSigma}
\safemath{\matOmega}{\bOmega}
\safemath{\matTheta}{\bTheta}

\safemath{\matidentity}{\matI}
\safemath{\matone}{\matO}

\safemath{\rnda}{A}
\safemath{\rndb}{B}
\safemath{\rndc}{C}
\safemath{\rndd}{D}
\safemath{\rnde}{E}
\safemath{\rndf}{F}
\safemath{\rndg}{G}
\safemath{\rndh}{H}
\safemath{\rndi}{I}
\safemath{\rndj}{J}
\safemath{\rndk}{K}
\safemath{\rndl}{L}
\safemath{\rndm}{M}
\safemath{\rndn}{N}
\safemath{\rndo}{O}
\safemath{\rndp}{P}
\safemath{\rndq}{Q}
\safemath{\rndr}{R}
\safemath{\rnds}{S}
\safemath{\rndt}{T}
\safemath{\rndu}{U}
\safemath{\rndv}{V}
\safemath{\rndw}{W}
\safemath{\rndx}{X}
\safemath{\rndy}{Y}
\safemath{\rndz}{Z}

\safemath{\rveca}{\bimA}
\safemath{\rvecb}{\bimB}
\safemath{\rvecc}{\bimC}
\safemath{\rvecd}{\bimD}
\safemath{\rvece}{\bimE}
\safemath{\rvecf}{\bimF}
\safemath{\rvecg}{\bimG}
\safemath{\rvech}{\bimH}
\safemath{\rveci}{\bimI}
\safemath{\rvecj}{\bimJ}
\safemath{\rveck}{\bimK}
\safemath{\rvecl}{\bimL}
\safemath{\rvecm}{\bimM}
\safemath{\rvecn}{\bimN}
\safemath{\rveco}{\bomO}
\safemath{\rvecp}{\bimP}
\safemath{\rvecq}{\bimQ}
\safemath{\rvecr}{\bimR}
\safemath{\rvecs}{\bimS}
\safemath{\rvect}{\bimT}
\safemath{\rvecu}{\bimU}
\safemath{\rvecv}{\bimV}
\safemath{\rvecw}{\bimW}
\safemath{\rvecx}{\bimX}
\safemath{\rvecy}{\bimY}
\safemath{\rvecz}{\bimZ}

\safemath{\rvecxi}{\bmxi}
\safemath{\rveclambda}{\bmlambda}
\safemath{\rvecmu}{\bmmu}
\safemath{\rvectheta}{\bmtheta}
\safemath{\rvecphi}{\bmphi}

\safemath{\rmatA}{\bimA}
\safemath{\rmatB}{\bimB}
\safemath{\rmatC}{\bimC}
\safemath{\rmatD}{\bimD}
\safemath{\rmatE}{\bimE}
\safemath{\rmatF}{\bimF}
\safemath{\rmatG}{\bimG}
\safemath{\rmatH}{\bimH}
\safemath{\rmatI}{\bimI}
\safemath{\rmatJ}{\bimJ}
\safemath{\rmatK}{\bimK}
\safemath{\rmatL}{\bimL}
\safemath{\rmatM}{\bimM}
\safemath{\rmatN}{\bimN}
\safemath{\rmatO}{\bimO}
\safemath{\rmatP}{\bimP}
\safemath{\rmatQ}{\bimQ}
\safemath{\rmatR}{\bimR}
\safemath{\rmatS}{\bimS}
\safemath{\rmatT}{\bimT}
\safemath{\rmatU}{\bimU}
\safemath{\rmatV}{\bimV}
\safemath{\rmatW}{\bimW}
\safemath{\rmatX}{\bimX}
\safemath{\rmatY}{\bimY}
\safemath{\rmatZ}{\bimZ}

\safemath{\rmatDelta}{\bimDelta}
\safemath{\rmatLambda}{\bimLambda}
\safemath{\rmatPhi}{\bimPhi}
\safemath{\rmatSigma}{\bimSigma}
\safemath{\rmatOmega}{\bimOmega}
\safemath{\rmatTheta}{\bimTheta}

\usepackage{amssymb}
\usepackage{amsfonts}
\usepackage{mathrsfs}
\usepackage{xspace}
\usepackage{bm}
\usepackage{fancyref}
\usepackage{textcomp}

\usepackage{multirow}
\usepackage{stmaryrd}

\newenvironment{textbmatrix}{	\setlength{\arraycolsep}{2.5pt}%
								\big[\begin{matrix}}{\end{matrix}\big]%
								\raisebox{0.08ex}{\vphantom{M}}}

\def\be{\begin{equation}}
\def\ee{\end{equation}}
\def\een{\nonumber \end{equation}}
\def\mat{\begin{bmatrix}}
\def\emat{\end{bmatrix}}
\def\btm{\begin{textbmatrix}}
\def\etm{\end{textbmatrix}}

\def\ba#1\ea{\begin{align}#1\end{align}}
\def\bas#1\eas{\begin{align*}#1\end{align*}}
\def\bs#1\es{\begin{split}#1\end{split}} 
\def\bg#1\eg{\begin{gather}#1\end{gather}}
\def\bml#1\eml{\begin{multline}#1\end{multline}}
\def\bi#1\ei{\begin{itemize}#1\end{itemize}}

\newcommand{\lefto}{\mathopen{}\left}

\DeclareMathOperator{\tr}{tr}				
\DeclareMathOperator{\Exop}{\opE}			

\newcommand{\Ex}[2]{\ensuremath{\Exop_{#1}\lefto[#2\right]}} 	

\safemath{\dirac}{\delta}					
\safemath{\krond}{\dirac}					

\safemath{\upto}{\uparrow}
\safemath{\downto}{\downarrow}
\safemath{\iu}{j}							
\safemath{\ev}{\lambda}						
\safemath{\hilseqspace}{l^{2}}				
\newcommand{\banachfunspace}[1]{\setL^{#1}}	
\safemath{\hilfunspace}{\banachfunspace{2}}	

\safemath{\SNR}{\textsf{SNR}} 				
\safemath{\PAR}{\textsf{PAR}} 				
\safemath{\No}{N_0}							
\safemath{\Es}{E_s}							
\safemath{\Eb}{E_b}							
\safemath{\EbNo}{\frac{\Eb}{\No}}
\safemath{\EsNo}{\frac{\Es}{\No}}

\DeclareMathOperator{\CHop}{\ensuremath{\opH}} 
\safemath{\tvir}{\rndh_{\CHop}}				
\safemath{\tvtf}{\rndl_{\CHop}}				
\safemath{\spf}{\rnds_{\CHop}}				
\safemath{\bff}{H_{\CHop}}					

\safemath{\ircf}{r_{h}}						
\safemath{\tftvcf}{r_{s}}					
\safemath{\tfcf}{r_{l}}						
\safemath{\bfcf}{r_{H}}						

\safemath{\tcorr}{c_h}						
\safemath{\scf}{c_{s}}						
\safemath{\tfcorr}{c_{l}}					
\safemath{\fcorr}{c_{H}}						

\safemath{\mi}{I}							
\safemath{\capacity}{C}						

\safemath{\normal}{\mathcal{N}}			
\safemath{\jpg}{\mathcal{CN}}			
\safemath{\mchain}{\leftrightarrow}		

\safemath{\dB}{\,\mathrm{dB}}
\safemath{\dBm}{\,\mathrm{dBm}}
\safemath{\Hz}{\,\mathrm{Hz}}
\safemath{\kHz}{\,\mathrm{kHz}}
\safemath{\MHz}{\,\mathrm{MHz}}
\safemath{\GHz}{\,\mathrm{GHz}}
\safemath{\s}{\,\mathrm{s}}
\safemath{\ms}{\,\mathrm{ms}}
\safemath{\mus}{\,\mathrm{\text{\textmu}s}}
\safemath{\ns}{\,\mathrm{ns}}
\safemath{\ps}{\,\mathrm{ps}}
\safemath{\meter}{\,\mathrm{m}}
\safemath{\mm}{\,\mathrm{mm}}
\safemath{\cm}{\,\mathrm{cm}}
\safemath{\m}{\,\mathrm{m}}
\safemath{\W}{\,\mathrm{W}}
\safemath{\mW}{\, \mathrm{mW}}
\safemath{\J}{\,\mathrm{J}}
\safemath{\K}{\,\mathrm{K}}
\safemath{\bit}{\,\mathrm{bit}}
\safemath{\nat}{\,\mathrm{nat}}

\safemath{\define}{\triangleq}			

\safemath{\equivalent}{\sim}
\safemath{\distas}{\sim}					
\safemath{\sdiff}{\Delta}				

\safemath{\reals}{\mathbb{R}}
\safemath{\positivereals}{\reals_{+}}
\safemath{\integers}{\mathbb{Z}}
\safemath{\posint}{\integers_{+}}
\safemath{\naturals}{\mathbb{N}}
\safemath{\posnaturals}{\naturals_{+}}
\safemath{\complexset}{\mathbb{C}}
\safemath{\rationals}{\mathbb{Q}}

\newcommand*{\fancyrefapplabelprefix}{app}		
\newcommand*{\fancyrefthmlabelprefix}{thm}		
\newcommand*{\fancyreflemlabelprefix}{lem}		
\newcommand*{\fancyrefcorlabelprefix}{cor}		
\newcommand*{\fancyrefdeflabelprefix}{def}		
\newcommand*{\fancyrefproplabelprefix}{prop}	
\newcommand*{\fancyrefobslabelprefix}{obs}		
\newcommand*{\fancyrefalglabelprefix}{alg}		
\newcommand*{\fancyrefremlabelprefix}{rem}		
\newcommand*{\fancyrefasmlabelprefix}{asm}	    

\frefformat{vario}{\fancyrefseclabelprefix}{Section~#1}
\frefformat{vario}{\fancyrefthmlabelprefix}{Theorem~#1}
\frefformat{vario}{\fancyreflemlabelprefix}{Lemma~#1}
\frefformat{vario}{\fancyrefcorlabelprefix}{Corollary~#1}
\frefformat{vario}{\fancyrefdeflabelprefix}{Definition~#1}
\frefformat{vario}{\fancyrefobslabelprefix}{Observation~#1}
\frefformat{vario}{\fancyrefasmlabelprefix}{Assumption~#1}
\frefformat{vario}{\fancyreffiglabelprefix}{Fig.~#1}
\frefformat{vario}{\fancyrefapplabelprefix}{Appendix~#1} 
\frefformat{vario}{\fancyrefproplabelprefix}{Proposition~#1}
\frefformat{vario}{\fancyrefalglabelprefix}{Algorithm~#1}
\frefformat{vario}{\fancyrefremlabelprefix}{Remark~#1}
\frefformat{vario}{\fancyrefeqlabelprefix}{(#1)}

\newtheorem{thm}{Theorem}

\newtheorem{lem}[thm]{Lemma} 

\safemath{\dictab}{[\,\dicta\,\,\dictb\,]}

\safemath{\ysig}{\bmy}
\safemath{\ysighat}{\hat{\ysig}}
\safemath{\ysigdim}{M}
\safemath{\xsig}{\bmx}
\safemath{\xsigdim}{N}
\safemath{\nx}{n_x}
\safemath{\zsig}{\bmz}
\safemath{\zsigdim}{\ysigdim}
\safemath{\rsig}{\bmr}
\safemath{\Adict}{\bA}
\safemath{\Adicttilde}{\widetilde{\Adict}}
\safemath{\Adictdim}{\outputdim\times\xsigdim}
\safemath{\avec}{\bma}
\safemath{\avectilde}{\tilde{\avec}}
\safemath{\Bdict}{\bB}
\safemath{\Bdicttilde}{\widetilde{\Bdict}}
\safemath{\Cdict}{\bC}
\safemath{\cvec}{\bmc}
\safemath{\Ddict}{\bD}
\safemath{\Ddictdim}{\ysigdim\times\xsigdim}
\safemath{\dvec}{\bmd}
\safemath{\Ddicttilde}{\widetilde{\bD}}
\safemath{\Bonb}{\bB}
\safemath{\bvec}{\bmb}
\safemath{\Bonbdim}{\ysigdim\times\ysigdim}
\safemath{\noise}{\bmn}
\safemath{\noisedim}{\ysigim}
\safemath{\err}{\bme}
\safemath{\errdim}{\ysigdim}
\safemath{\errset}{\setE}
\safemath{\nerr}{n_e}
\safemath{\delop}{\bP_\errset}
\safemath{\delopc}{\bP_{{\errset}^c}}

\safemath{\cplxi}{\imath}
\safemath{\cplxj}{\jmath}

\safemath{\dict}{\matD}
\safemath{\inputdim}{N}		
\safemath{\outputdim}{M}		
\safemath{\sparsity}{S}	
\safemath{\inputdimA}{{N_a}}	
\safemath{\inputdimB}{{N_b}}	
\safemath{\elemA}{{n_a}}	
\safemath{\elemB}{{n_b}}	
\safemath{\resA}{\matR_a}	
\safemath{\resB}{\matR_b}	
\safemath{\subD}{\matS} 
\safemath{\subA}{\matS_a} 
\safemath{\subB}{\matS_b} 
\safemath{\dicta}{\matA} 	
\safemath{\dictb}{\matB} 	
\safemath{\hollowS}{H}
\safemath{\hollowA}{H_a}
\safemath{\hollowB}{H_b}
\safemath{\cross}{Z}
\safemath{\coh}{\mu_d}			
\safemath{\coha}{\mu_a}			
\safemath{\cohb}{\mu_b}			
\safemath{\mubs}{\nu}	
\safemath{\cohm}{\mu_m} 
\safemath{\dictset}{\setD}	
\safemath{\dictsetp}{\dictset(\coh,\coha,\cohb)}	
\safemath{\dictsetgen}{\dictset_\text{gen}}
\safemath{\dictsetgenp}{\dictsetgen(\coh)}
\safemath{\dictsetonb}{\dictset_\text{onb}}
\safemath{\dictsetonbp}{\dictsetonb(\coh)}

\safemath{\leftside}{U}
\safemath{\rightsideA}{R_a}
\safemath{\rightsideB}{R_b}

\safemath{\indexS}{\setI_S} 

\safemath{\na}{n_a}			
\safemath{\nb}{n_b}			
\safemath{\coeffa}{p_i}	
\safemath{\coeffb}{q_j}	
\safemath{\seta}{\setP}		
\safemath{\setb}{\setQ}     
\safemath{\setw}{\setW}	
\safemath{\setz}{\setZ}	
\safemath{\cola}{\veca}		
\safemath{\colb}{\vecb}		
\safemath{\cold}{\vecd}		
\safemath{\inputvec}{\vecx} 	
\safemath{\error}{\vece}	
\safemath{\noiseout}{\vecz} 	
\safemath{\inputvecel}{x}
\safemath{\inputveca}{\vecx_a}
\safemath{\inputvecb}{\vecx_b}
\safemath{\outputvec}{\vecy}	
\safemath{\lambdamin}{\lambda_{\mathrm{min}}}

\safemath{\elltwo}{\ell_2}
\safemath{\ellone}{\ell_1}
\safemath{\ellzero}{\ell_0}
\safemath{\ellinf}{\ell_\infty}
\safemath{\ellinftilde}{\ell_{\widetilde\infty}}
\safemath{\licard}{Z(\coh,\coha,\cohb)}
\safemath{\xsol}{\hat{x}}
\safemath{\xbord}{x_b}		
\safemath{\xstat}{x_s}		
\safemath{\xstatLone}{\tilde{x}_s}
\safemath{\order}{\mathcal{O}} 
\safemath{\scales}{\Theta} 
\safemath{\ones}{\mathbf{1}} 
\safemath{\zeroes}{\mathbf{0}} 
\safemath{\thlone}{\kappa(\coh,\cohb)} 
\safemath{\constoneA}{\delta} 
\safemath{\constoneB}{\epsilon} 
\safemath{\nlarge}{L}				   
\safemath{\sumlarge}{S_\nlarge}
\safemath{\maxlarger}{P_\nlarge}	   
\safemath{\Pzero}{\textrm{P0}}	
\safemath{\Pone}{\textrm{P1}}
\safemath{\vecfir}{\vecw}			 
\safemath{\vecsec}{\vecz}
\safemath{\elvecfir}{w}              
\safemath{\elvecsec}{z}				 
\safemath{\nlargefir}{n}
\safemath{\normout}{\gamma}
\safemath{\auxfun}{h}
\safemath{\supp}{\textrm{supp}}

\safemath{\indexa}{\ell}
\safemath{\indexb}{r}
\safemath{\indexc}{i}
\safemath{\indexd}{j}

\safemath{\project}{P}

\allowdisplaybreaks

\IEEEoverridecommandlockouts

\begin{document}
\title{Feedforward Architectures for Decentralized Precoding in Massive MU-MIMO Systems}
\author{\IEEEauthorblockN{Kaipeng Li$^\text{1}$, Charles Jeon$^\text{2}$, Joseph R. Cavallaro$^\text{1}$, and Christoph Studer$^\text{2}$} \\
\IEEEauthorblockA{$^\text{1}$Department of Electrical and Computer Engineering, Rice University, Houston, TX}
\IEEEauthorblockA{$^\text{2}$School of Electrical and Computer Engineering, Cornell University, Ithaca, NY} 
\thanks{This work was supported in part by Xilinx, Inc., the US NSF under grants CNS-1265332, ECCS-1232274, ECCS-1408370, CNS-1717218, ECCS-1408006, CCF-1535897,  CCF-1652065, CNS-1717559, and with hardware and software support from the Texas Advanced Computing Center and the Nvidia Technology Center (the PSG Cluster) with DGX-1 multi-GPU systems.}
}

\maketitle


\vspace{-1cm}
\begin{abstract}

Massive multi-user multiple-input multiple-output (MU-MIMO) enables significant gains in spectral efficiency and link reliability compared to conventional small-scale MIMO technology.
Furthermore, linear precoders, e.g., using zero forcing or Wiener filter (WF) precoding, are sufficient to achieve excellent error-rate performance in the massive MU-MIMO downlink. 
However, these methods necessitate centralized processing at the base-station (BS), which causes (i) excessively high interconnect and chip input/output data rates, and (ii) high implementation complexity. 
We propose two feedforward architectures and corresponding decentralized WF precoders that parallelize precoding across multiple computing fabrics, effectively mitigating the issues of centralized approaches. 
To demonstrate the efficacy of our decentralized precoders, we provide implementation results on  a multi-GPU system, which show that our solutions achieve throughputs in the Gbit/s regime while achieving (near-)optimal error-rate performance in the massive MU-MIMO downlink.
\end{abstract}


\section{Introduction}
Massive multi-user (MU) multiple-input multiple-output (MIMO) will be among the core technologies of fifth-generation (5G) cellular wireless systems~\cite{5gbe}.
The key idea of this technology is to equip the infrastructure base-stations (BSs) with hundreds to thousands of antenna elements while serving tens of user equipments (UEs) at the same time and in the same frequency band. 
The fine-grained nature of beamforming enabled by massive MU-MIMO antenna arrays and coherent transmission yields significantly improved spectral efficiency, coverage, and range compared to that of traditional, small-scale multi-antenna wireless systems~\cite{JH_13, mimo_overview}. 
Unfortunately, the advantages of massive MU-MIMO come at the cost of significant practical implementation challenges, which must be solved to realize the gains of this technology in practice.

\subsection{Interconnect Bandwidth and Complexity Bottlenecks}
As discussed in~\cite{puglielli2015scalable,Li_JETCAS17, quantizeMM, erikdbp}, the excessively high amount of data that must be transferred between the baseband processing unit and the antenna array is among the most critical challenges. 
For example, the raw baseband data rates, from and to the radio-frequency (RF) chains, of a $128$ antenna massive MU-MIMO system with $10$\,bit digital-to-analog converters (DACs) for a bandwidth of $40$\,MHz exceed $200$\,Gbit/s, which not only poses significant challenges for existing interconnect technology, such as the Common Public Radio Interface (CPRI)~\cite{cpri}, but also for the input/output (I/O) interfaces of existing computing fabrics, such as application-specific integrated circuits (ASICs), field-programable gate arrays (FPGAs), or graphics processing units (GPUs). 
While maximum ratio transmission (MRT) enables fully distributed channel estimation (CHEST) and beamforming in the   downlink (BS transmits to the UEs), which alleviates the bandwidth and I/O bottlenecks, MRT is unable to fully exploit the spectral-efficiency advantages of massive MU-MIMO~\cite{JH_13}.  
More sophisticated precoding strategies, such as zero-forcing (ZF) or Wiener filter (WF) precoding~\cite{wfprecoding}, enable near-optimal spectral efficiency. However, such precoding methods require centralized baseband processing which results in high interconnect bandwidth,  I/O data rates, and  complexity~\cite{JH_13}. 

\subsection{Decentralized Baseband Processing}
To mitigate the bandwidth and complexity bottlenecks of centralized baseband processing algorithms, a variety of solutions have been proposed recently. For example, existing massive MU-MIMO testbeds parallelize the computations across orthogonal subcarriers~\cite{bigstation, lund}. Parallelization across subcarriers, however, exhibits dependence between the subcarriers and all antenna elements, which prevents its straightforward use for arrays with thousands of antennas.
More recently, the papers~\cite{ Li_Asilomar_16, Li_GlobalSIP_16} proposed \emph{decentralized baseband processing} (DBP), an approach that decentralizes the key  computations required for baseband processing in massive MU-MIMO systems in order to alleviate the bandwidth and complexity bottlenecks. The idea of DBP is to partition the antenna array into clusters, each associated with separate RF circuitry and processing fabrics. Each cluster then only communicates with the associated processing fabrics that carry out (de-)modulation, channel estimation, data detection, and precoding, while only exchanging a small amount of consensus information among the clusters. 
However, as it has been demonstrated with real-world implementations on GPU clusters  in~\cite{Li_JETCAS17}, the exchange of consensus information among the clusters negatively affects the processing latency and throughput. 
As an effective remedy, reference \cite{isit17} proposed the use of \emph{feedforward architectures} for equalization the uplink (UEs transmit to BS). Such architectures have, up to this point, not been studied for the downlink. 

\subsection{Contributions}
We  propose two new feedforward architectures and corresponding algorithms for decentralized precoding in the massive MU-MIMO downlink. Both architectures are compatible with the ones proposed for the uplink in~\cite{isit17} and prevent the repeated exchange of consensus information to effectively reduce latency and throughput. For both architectures, we propose linear precoding algorithms that build upon the WF precoder, which minimizes the mean-square error (MSE) at the UE side. 
We show that the WF precoder for the \emph{partially decentralized} (PD) architecture achieves the same performance as the centralized WF precoder; the WF precoder for the \emph{fully decentralized} (FD) architecture further reduces the interconnect bandwidth at a small error-rate performance loss.
We demonstrate the efficacy of our feedforward architectures and precoding algorithms using real-world implementations on a multi graphics processing unit (GPU) system.
Our implementation results reveal that decentralized precoding with feedforward architectures reaches throughputs in the Gb/s regime on a multi-GPU system while achieving (near-)optimal error-rate performance.

\subsection{Notation}
Lowercase and uppercase boldface letters denote column vectors and matrices, respectively. 
The transpose and Hermitian transpose of the matrix~$\bA$ are dentoed by~$\bA^T$ and~$\bA^H$, respectively.
The $M\times M$ identity matrix is denoted by $\bI_M$.
We use~$\tr(\bA)$ to denote the trace of the matrix $\bA$ and~$\Ex{\bma}{\cdot}$ to denote expectation with respect to the random vector~$\bma$.

\begin{figure*}[t]
\centering
\subfigure[Partially decentralized (PD) precoding architecture.]{\includegraphics[height=0.5\columnwidth]{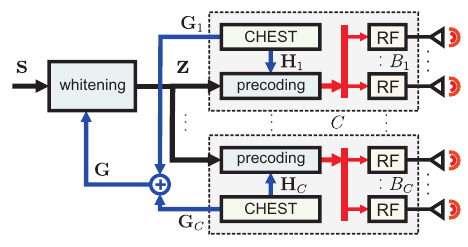}
\label{fig:pda}
}
\hspace{0.5cm}
\subfigure[Fully decentralized (FD) precoding architecture.]{\includegraphics[height=0.5\columnwidth]{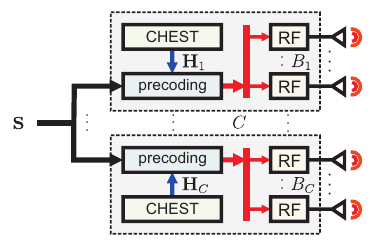}
\label{fig:fda}
}
\vspace{0.1cm}
\caption{Partially decentralized (PD) and fully decentralized (FD) precoding architectures for the massive MU-MIMO downlink with $C$ clusters.
(a) PD performs decentralized channel estimation (CHEST) in the uplink and averages the partial Gram matrices $\bG_c$ while feeding them to the centralized whitening unit; the $\oplus$ operator denotes  matrix additions.
In the downlink, precoding is performed in two steps: centralized whitening followed by decentralized precoding in each cluster. 
(b) FD performs decentralized CHEST in the uplink. In the downlink, precoding is performed locally at each cluster in a fully decentralized manner.}
\label{fig:architectures}
\end{figure*}

\section{System Model and Centralized Precoding}

We now introduce the system model and discuss the basics of centralized precoding for massive MU-MIMO systems.

\subsection{Downlink System Model}
We focus on the massive MU-MIMO downlink. The system consists of a base-station (BS) with~$B$ antennas serving~$U$ single-antenna user-equipments (UEs) in the same time-frequency resource. 
We consider a block-fading and narrowband\footnote{An extension to wideband systems that use orthogonal frequency division multiplexing (OFDM) is straightforward and considered in \fref{sec:multigpu}.} scenario modeled as follows:
\begin{align} \label{eq:downlinkmodel}
\bmy[k] = \bH \bmx[k] + \bmn[k], \quad k=1,\ldots,K.
\end{align}
Here, the $U$-dimensional vector $\bmy[k]=\big[\, y_1[k],\ldots,y_U[k]\big]^T$ contains the signals received at all $U$ UEs with $y_u[k]\in\complexset$ corresponding to the signal received at UE $u$ in time slot~$k$. 
The matrix $\bH\in\complexset^{U\times B}$ represents the downlink MIMO channel and is assumed to remain constant for $K$ time slots.
The vector $\bmn[k]\in\complexset^U$ models additive noise and is assumed to be i.i.d.\ circularly-symmetric complex Gaussian with variance~$\No$ per complex entry. We assume the channel matrix~$\bH$ and noise variance $\No$ to be known perfectly at the BS. 
The precoded vector $\bmx[k]\in\complexset^B$ at time slot $k$  is given by the function 
\begin{align*}
\bmx[k]=\setP(\bms[k],\bH,\No,\rho^2),
\end{align*} 
which depends on transmit signal vector $\bms[k]\in\setO^U$, where $\setO$ is the constellation set (e.g., 64-QAM), the  channel matrix $\bH$, the noise variance $\No$, and the power constraint $\rho^2$. 
The precoded vector is assumed to satisfy the average power constraint
\begin{align}  \label{eq:powerconstraint}
\Ex{\bms}{\|\bmx[k]\|^2}\leq \rho^2, \quad  k=1,\ldots,K,
\end{align}
and the vector $\bms[k]=\big[s_1[k],\ldots,s_U[k]\big]^T$ contains the information symbols $s_u[k]\in\setO$ to be transmitted to UE $u$ in time slot~$k$. 
In what follows, we omit the time-slot index $k$.

\subsection{Linear Wiener Filter Precoding}
\label{sec:centralizedprecoding}
Since the UEs are unable to perform joint signal processing, the BS has to precode the information symbols with the goal of mitigating multi-user interference (MUI).
The literature describes numerous optimization criteria for precoding, such as sum-rate throughput or error probability~\cite{precodingsurvey}. In what follows, we focus on linear precoders of the form $\bmx=\bP\bms$ that minimize the mean-square error (MSE) between the estimated symbol vector $\hat\bms$ and the transmit signal vector $\bms$.
Since coherent transmission with a multi-antenna BS leads to an array gain, we assume that the UEs are able to scales the received signals~$y_u$, $u=1,\ldots,U$, by a precoding factor $\beta_u\in\complexset$, i.e., the UEs compute symbol estimates as follows:
\begin{align*}
\hat s_u = \beta_u y_u.
\end{align*}
While  each UE $u$ would able to estimate their own precoding factor $\beta_u$, we design precoders that minimize the MSE for a joint\footnote{Designing precoders for the case of having individual precoding factors $\beta_u$, $u=1,\ldots,U$, is challenging and left for future work.} precoding factor $\beta\in\complexset$ defined as~\cite{wfprecoding}
\begin{align*}
\textit{MSE} & = \Ex{\bms,\bmn}{\|\bms-\hat\bms\|^2} = \Ex{\bms,\bmn}{\|\bms-\beta\bmy\|^2} \\
& =  \Ex{\bms}{\|\bms-\beta\bH\bmx\|^2} + |\beta|^2 U \No.
\end{align*}
The resulting  MSE-optimal linear precoding matrix $\bP\in\complexset^{B\times U}$ can be designed by solving the following optimization problem 
\begin{align}
\{\bP^\text{WD},\beta^\text{WF}\} = \left\{
\begin{array}{ll}
  \underset{\bP,\beta}{\text{minimize}}  & \Ex{\bms}{\|\bms-\beta\bH\bP\bms\|^2} + \beta^2 U \No \\
\text{subject to} & \Ex{\bms}{\|\bmx\|^2} \leq \rho^2. 
\end{array}\right. \notag \\[-0.55cm]
\label{sec:WFproblem}
\end{align}
The solution to this optimization problem is known as the Wiener filter (WF) precoder~\cite{wfprecoding} and is summarized by the following result; a short proof is given in \fref{app:WFprecoderproof}.
\begin{thm}\label{thm:WFprecoder}
The Wiener filter (WF) precoding matrix $\bP^\text{WF}$ is given by  $\bP^\text{WF} = \frac{1}{\beta^\text{WF}}\bQ^\text{WF}$, where we define the matrix
\begin{align} \label{eq:originalQmatrix}
\bQ^\text{WF} = \left(\bH^H\bH + \kappa^\text{WF} \bI_B \right)^{-1}\bH^H.
\end{align}
The associated regularization parameter $\kappa^\text{WF}$  and precoding factor~$\beta^\text{WF}$ are defined as 
\begin{align} \label{eq:precodingfactor}
\kappa^\text{WF} =  \frac{U\No}{\rho^2} \quad \text{and} \quad \beta^\text{WF}  = \sqrt{\frac{\tr(\bQ^H\bQ) \Es}{\rho^2}}.
\end{align}
\end{thm}

A straightforward computation of the precoding factor $\beta^\text{WF}$ in \fref{eq:precodingfactor} involves the inversion of a $B\times B$ matrix followed by a number of matrix-matrix multiplications. 
We propose a computationally-efficient alternative that can be implemented efficiently given the $U\times U$ \emph{whitening matrix}~$\bA^{-1}$ has been precomputed; a proof is given in \fref{app:precodingfactor}.
\begin{lem}\label{lem:precodingfactor}
The precoding factor $\beta^\text{WF}$ of the WF precoder in~\fref{eq:precodingfactor}  can  be computed efficiently as follows: 
\begin{align} \label{eq:beta}
\beta^\text{WF} = \sqrt{\frac{\Es}{\rho^2}\! \left(\tr\left(  \bA^{-1} \right) - \kappa^\text{WF}  \|\bA^{-1}\|_F^2 \right)}. 
\end{align}
\end{lem}

\section{Decentralized Precoding }

We next propose two decentralized precoding schemes that rely on feedforward architectures and linear WF precoding. 
 
\subsection{System Model for Decentralized Precoding}
We now extend the feedforward architecture put forward in~\cite{isit17} for the uplink by the capability to perform downlink precoding. 
To this end, we partition the BS antenna array into $C\geq1$ clusters, each associated with $B_c=w_cB\in\mathbb{N}^+$ BS antennas so that $w_c\in (0,1]$ and $\sum_{c=1}^C w_c = 1$. Each cluster contains local RF circuitry and requires access to only local channel state information (CSI) acquired in the uplink via reciprocity. 
By omitting the time-slot index $k$, we can rewrite the downlink system model in \fref{eq:downlinkmodel} as
 \begin{align} \label{eq:decentralizeddownlinkmodel}
\bmy = \sum_{c=1}^C \bH_c \bmx_c + \bmn, 
\end{align}
where $\bH=\big[\bH_1,\ldots,\bH_C\big]$ with $\bH_c=\complexset^{U\times B_c}$ and $\bmx^T=\big[\bmx_1^T,\ldots,\bmx_C^T\big]$ with $\bmx_c\in\complexset^{B_c}$, we see that each cluster $c=1,\ldots,C$ has to generate a precoding vector $\bmx_c$ with information of the per-cluster channel matrix $\bH_c$, the noise variance $\No$, the power constraint $\rho^2$, and the transmit symbol vector $\bms$, i.e., the precoding functions are as follows:
\begin{align} \label{eq:precodingfunctiondecentralized}
\bmx_c=\setP_c(\bms,\bH_c,\No,\rho^2), \quad c=1,\ldots,C.
\end{align}
Since each of these functions only depends on local CSI contained in $\bH_c$, the exchange of CSI is reduced significantly---the vector $\bms$ is the only signal that must be broadcast to all clusters.
We now present two distinct feedforward architectures that perform decentralized precoding, differing in the amount of CSI that must be exchanged during the training phase.
  
\subsection{Partially-Decentralized WF Precoding}
The first feedforward architecture  is  illustrated in \fref{fig:pda} and called \emph{partially decentralized WF (PD-WF) architecture}. The operating principle can be derived directly from \fref{eq:WFprecoderinvertedcompact}, which results in the precoding rule
\begin{align*}
\bmx = \frac{1}{\beta^\text{WF}}\bH^H \bA^{-1}\bms.
\end{align*}
The idea of PD-WF precoding is to first whiten the transmit vector $\bms$ at a centralized \emph{whitening node} as follows:
\begin{align*}
\bmz = \frac{1}{\beta^\text{WF}}\bA^{-1}\bms.
\end{align*}
The whitened transmit vector $\bmz$ is then transmitted to each cluster, which independently compute $\bmx_c=\bH^H_c\bmz$. 

Clearly, this PD-WF architecture requires the whitening matrix~$\bA^{-1}$ as well as the precoding factor $\beta^\text{WF}$ to be calculated at the centralized whitening node---both of these quantities require the computation of the Gram matrix $\bG$.
To compute this matrix in an decentralized architecture, we follow the approach for PD equalization put forward in~\cite{wfprecoding}, where each cluster $c=1,\ldots,C$, first computes the local Gram matrix $\bG_c=\bH_c\bH_c^H$ after estimating the channel in the uplink phase, followed by computing the (centralized) Gram matrix $\bG=\sum_{c=1}^C\bG_c$ in a feedforward adder tree; see the blue feedback path in \fref{fig:pda}. The centralized whitening node then computes the whitening matrix~$\bA^{-1}$ and the precoding factor~$\beta^\text{WF}$ as detailed in \fref{sec:centralizedprecoding}.
Since we have that
\begin{align*}  
\sum_{c=1}^C \bH_c \bmx_c  = \sum_{c=1}^C \bH_c \bH^H_c \bA^{-1}\bms  = \bG \bA^{-1}\bms = \bH\bP^\text{WF} \bms,
\end{align*}
the PD-WF architecture implements exactly the centralized WF precoder from \fref{thm:WFprecoder} but in a decentralized fashion.

\newcommand{\spacers}{0.55}
\begin{figure*}
\centering
\begin{tabular}{cp{0.2cm}cp{0.2cm}c}
(a) $B=256$, $B_c=128$, $C=2$ & & (b) $B=256$, $B_c=64$, $C=4$ & & (c) $B=256$, $B_c=32$, $C=8$  \\
\includegraphics[width=\spacers\columnwidth]{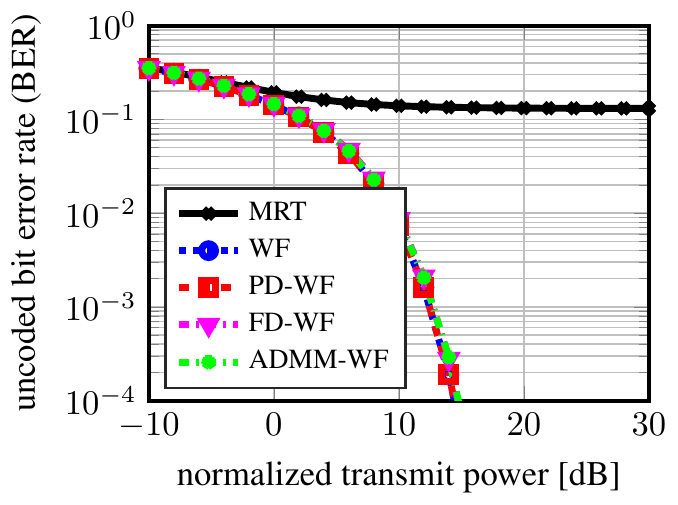} 
& & \includegraphics[width=\spacers\columnwidth]{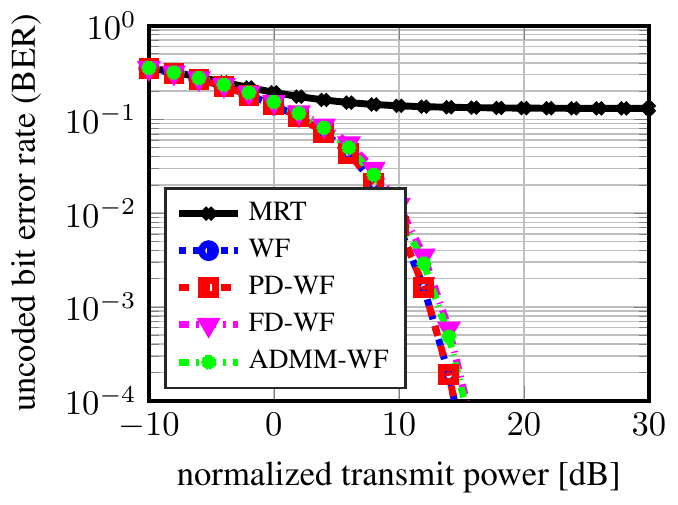} 
& & \includegraphics[width=\spacers\columnwidth]{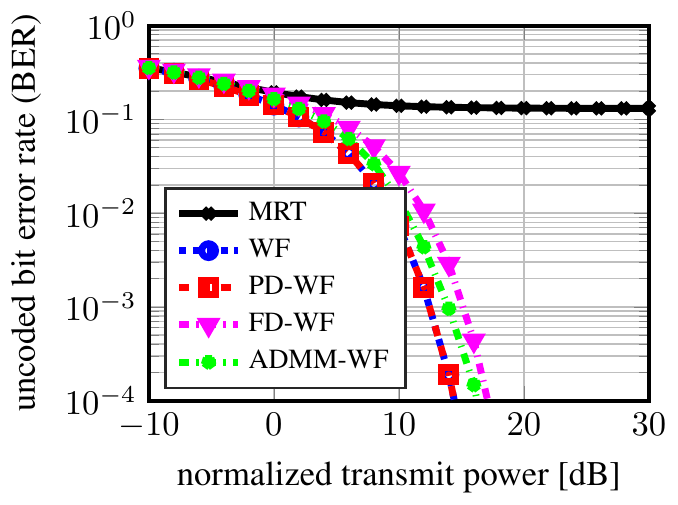} \\
PD-WF:  0.879\,ms,  0.979\,Gb/s & & PD-WF: 0.678\,ms / 1.270\,Gb/s & & PD-WF: 0.607\,ms / 1.418\,Gb/s\tabularnewline
FD-WF: 0.789\,ms,  1.091\,Gb/s & & FD-WF: 0.571\,ms / 1.507\,Gb/s & & FD-WF: 0.472\,ms / 1.824\,Gb/s \\[0.4cm]
 (d) $B=64$, $B_c=32$, $C=2$ & & (e) $B=128$, $B_c=32$, $C=4$ & & (f) $B=256$, $B_c=32$, $C=8$  \\
\includegraphics[width=\spacers\columnwidth]{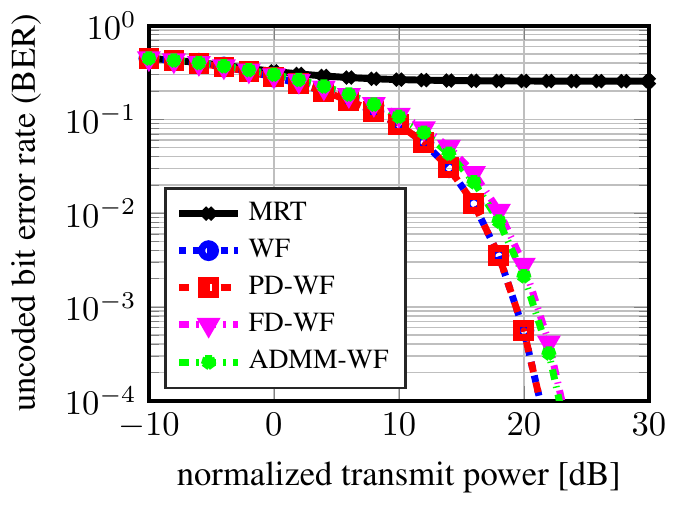} 
& & \includegraphics[width=\spacers\columnwidth]{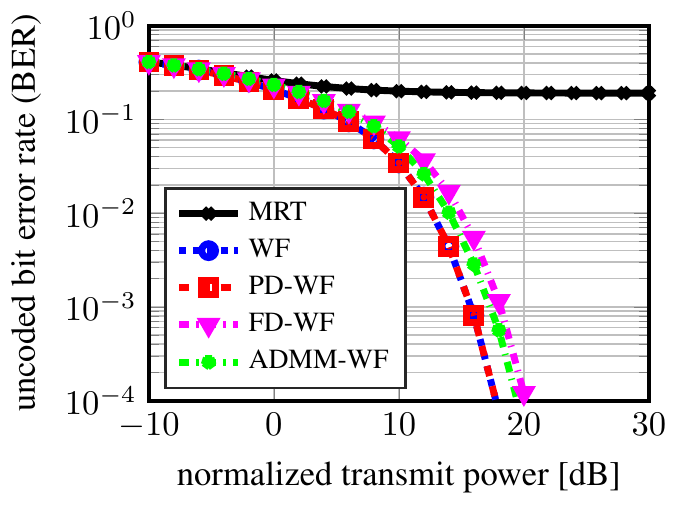} 
& & \includegraphics[width=\spacers\columnwidth]{SIM/paperplots_wide/ERR_16x256_C8_pilot_64QAM_1000Trials_0.pdf} \\
PD-WF:  0.532\,ms / 1.618\,Gbps & & PD-WF:  0.559\,ms / 1.540\,Gbps  & & PD-WF:  0.607\,ms / 1.418\,Gbps\tabularnewline
FD-WF: 0.441\,ms / 1.952\,Gbps & & FD-WF: 0.451\,ms / 1.909\,Gbps & & FD-WF: 0.472\,ms / 1.824\,Gbps \\[0.15cm]
\end{tabular}
\caption{Uncoded bit error-rate, latency,  and throughput  result for decentralized baseband processing with $U=16$ users. Top row: fixed number of BS antennas $B=256$, varying cluster size $B_c$ and number of clusters $C$. Bottom row: fixed cluster size $B_c=32$, varying number of BS antennas $B$ and number of clusters $C$. PD-WF achieves the same error-rate performance as centralized precoding;  FD-WF achieves near-WF performance for cluster sizes $B_c\geq32$; the ADMM-based WF method outperforms FD-WF but requires iterative exchange of consensus information resulting in higher latency.}
\label{fig:berandtp}
\end{figure*}

\subsection{Fully-Decentralized WF Precoding}
The second feedforward architecture, called  \emph{fully decentralized WF (FD-WF) architecture}, is illustrated in  \fref{fig:fda} and avoids transmitting partial Gram matrices to the centralized whitening node.
The key idea of this architecture is to first broadcast the transmit vector $\bms$ to each cluster, and then compute the local precoding vector as follows $\bmx_c=\bP_c\bms$. 
In order to adhere to the (total) power constraint in \fref{eq:powerconstraint}, we have to define a per-cluster power constraint $\Ex{}{\|\bmx_c\|^2} \leq 
\rho_c^2$ for which $\sum_{c=1}^C \rho^2_c = \rho^2$.
In what follows, we allocate the same amount of power\footnote{We investigated a number of strategies that allocate different power levels to each cluster. Such methods did not provide significant performance advantages in massive MU-MIMO, but may, for example, be critical for cell-free massive MU-MIMO systems in which the clusters are spread over a large area~\cite{ngo2017cell}.} to each cluster, i.e., $\rho_c^2=\rho^2/C$, which results in the following precoder carried out at each cluster
\begin{align*}
\bmx_c = \sqrt{\frac{\rho_c^2}{\tr(\bQ_c^H\bQ_c)\Es}} \bQ_c \bms.
\end{align*}
The remaining piece is to identify a suitable choice of the regularization parameters $\kappa_c$ that are used to calculate the matrices~$\bQ_c$.  
A straightforward way would be to assume that each cluster operates independently and to set the regularization parameter to $U\No/\rho^2_c$.
In practice, however, it turns out that this choice of the regularization parameter is overly pessimistic.
Since computing an optimal set of regularization parameters is difficult in the decentralized scenario, we simply set   
\begin{align} \label{eq:regularization}
\kappa_c =  \tau_c \frac{U\No}{\rho^2_c}, \quad c=1,\ldots,C,
\end{align}
and tune the parameters $\tau_c\in[0,\infty)$ for best error-rate performance via numerical simulations.  Specifically, we use
\begin{align*}
\bQ_c = \left\{
\begin{array}{ll}
\left(\bH^H_c\bH_c + \kappa_c \bI_{B_c} \right)^{-1}\bH^H_c & \text{if }B_c<U\\
\bH^H_c\left(\bH_c\bH^H_c + \kappa_c\bI_U \right)^{-1} & \text{if } B_c\geq U,
\end{array}\right.
\end{align*}
which further reduces the computational complexity depending on the number of antennas per cluster.

\subsection{Simulation Results}
\label{sec:ber}
We now show uncoded bit error-rate (BER) simulation results for a Rayleigh fading massive MU-MIMO system with 64-QAM. Figs.~\ref{fig:berandtp} (a), (b), (c)  show the BER for $B=256$ BS antennas, with varying cluster sizes $B_c=128,64,32$, and number of clusters $C=2,4,8$. Figs.~\ref{fig:berandtp} (d), (e), (f) show the BER for a fixed cluster size $B_c=32$, with a varying number of BS antennas  $B=64,128,256$, and number of clusters $C=2,4,8$. 
For each antenna configuration, we compare the performance of the proposed decentralized solutions PD-WF and  FD-WF, as well as existing methods including centralized WF precoding, fully-distributed MRT, and the fully-decentralized ADMM-based WF precoder from~\cite{Li_JETCAS17}.

Evidently, PD-WF achieves the same BER as the centralized WF precoder for all antenna configurations. In contrast, FD-WF suffers a moderate BER loss if $B_c$ is small. To minimize the performance loss of FD-WF precoding, we have tuned the parameter $\tau_c$ in \fref{eq:regularization}. Concretely, we found that  $\tau_c=0.125$ performs well for a broad range of antenna and cluster configurations; a corresponding theoretical analysis is left for future work.
In addition, we see that the fully decentralized ADMM-based WF precoder proposed in~\cite{Li_JETCAS17} is able to outperform FD-WF precoding but requires multiple iterations of consensus exchange  (we use two ADMM iterations) that dramatically reduces the throughput   due to the typically high interconnect latency between antenna clusters; see~\cite{Li_JETCAS17} for the details.

\section{Multi-GPU Implementation}
\label{sec:multigpu}
As a proof-of-concept, we now present a multi-GPU implementation  of PD-WF and FD-WF precoding, and show corresponding throughput and latency results. 

\subsection{System Architecture}
\begin{figure}[tp]
\centering
\includegraphics[width=0.9\columnwidth]{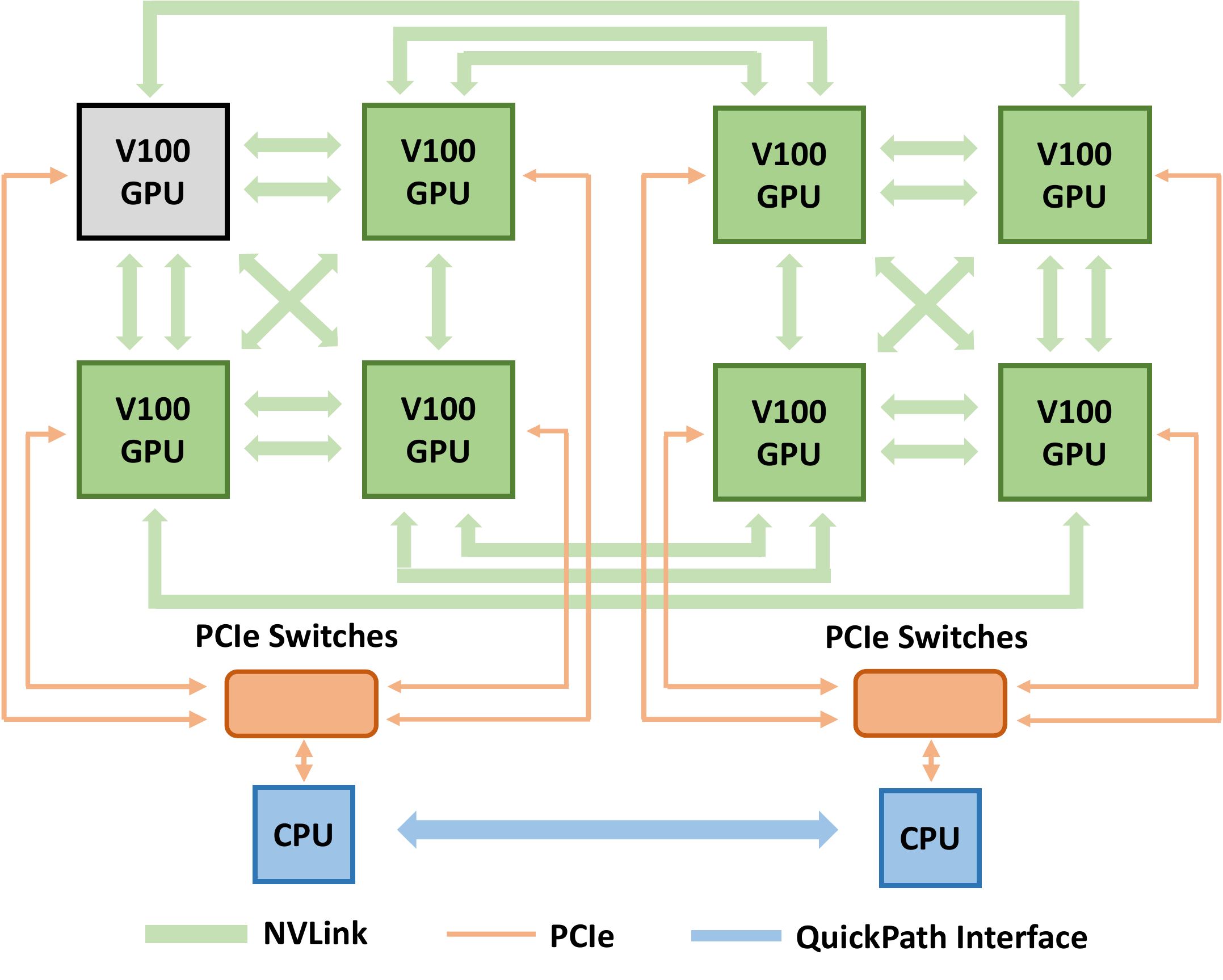}
\vspace{-0.2cm}
\caption{Overview of the experimental platform with up to eight Tesla Volta GPUs and high speed NvLink GPU-to-GPU interconnect~\cite{dgx}. The upper-left GPU is the master GPU that collects results from other GPUs, performs centralized computations for PD-WF, and broadcasts the transmit vectors $\mathbf{s}[k]$ for FD-WF or the whitened vectors $\mathbf{z}[k]$ for PD-WF  to other GPUs.} 
\label{fig:sys_arch}
\vspace{-0.3cm}
\end{figure}

We implemented PD-WF and FD-WF precoding on an Nvidia DGX-1 multi-GPU system~\cite{dgx}, as illustrated in \fref{fig:sys_arch}. 
The architecture consists of two $2$-core Intel Xeon E5-2698 v4 CPUs and eight Tesla V100 Volta GPUs with $300$\,GB/s bi-directional NvLink GPU-to-GPU communication links. 
Each Tesla V100 GPU contains $5120$ CUDA cores with 16\,GB high bandwidth memory~(HBM). 
For PD-WF and FD-WF precoding, we use the message passing interface (MPI) library OpenMPI to generate a total of $C$ processes in the multi-GPU system, where each process controls a GPU for accelerating the decentralized local workload using CUDA~\cite{cuda} with CUDA v9.1. 
While FD-WF only requires broadcasting of transmit signals $\mathbf{s}[k]$ across GPUs prior to the precoding computations, PD-WF necessitates gathering of the local Gram matrices from all GPUs via sum reduction at the centralized whitening unit (in the master GPU as shown in \fref{fig:sys_arch}), and broadcasting of the whitened vectors $\mathbf{z}[k]$. 
These message passing operations are implemented using the NVIDIA Collective Communications Library (NCCL)\cite{nccl} v2.1 that builds on MPI for high efficiency over NvLink.

\subsection{Implementation Details}
To increase the throughput on the multi-GPU system, we need to feed the GPU a sufficient amount of workloads to fully exploit the available resources. In what follows, we assume the downlink transmission with orthogonal frequency division multiplexing (OFDM) with $N_\text{sc}$ subcarriers. 
Each OFDM subcarrier corresponds to an independent narrowband block-fading downlink system as in \fref{eq:downlinkmodel}, where we assume that the  channel remains constant across $K$ OFDM symbols.
The vector $\mathbf{s}^w[k]$ indicates the transmit vector $\mathbf{s}[k]$ on subcarrier~$w$ in time slot $k$. In our implementations, we aggregate the precoding workloads for $K$ OFDM symbols, each including $N_\text{sc}$ subcarriers, and process them together in parallel to improve the GPU occupancy and throughput. In what follows, we omit the superscript $^w$ as well as the OFDM symbol index $k$. 

\subsubsection{PD-WF Implementation}
For PD-WF, we invoke~$C$ MPI processes that control $C$ GPUs, and each process initializes computation of the local Gram matrix $\mathbf{G}_c$ using the local channel $\mathbf{H}_c$ on a certain GPU. Within each GPU, we calculate $N_\text{sc}$ such $\mathbf{G}_c$ matrices to achieve high throughput.
These matrix multiplications can be efficiently implemented using the \emph{cuBLAS}~\cite{cublas} library; specifically, we use the  \texttt{cublasCgemmBatched} function for complex-valued floating-point arithmetic. Once all local $\mathbf{G}_c$ matrices have been computed, we gather $\mathbf{G}_c$ from all $C$ GPUs to a reduced sum (resulting in the global Gram matrix $\mathbf{G}$) at the master GPU using the NCCL \texttt{ncclReduce} function.
The NCCL library leverages \emph{CUDA-aware MPI}~\cite{cudaaware} for direct GPU-to-GPU memory copy over high-speed NvLink interconnect.

We compute $\mathbf{A}=\mathbf{G}+\kappa^{\text{WF}}\mathbf{I}_U$ for $N_\text{sc}$ 
subcarriers (corresponding to a given OFDM symbol $k$)
at the master GPU in parallel. We then invert this matrix using the \emph{cuBLAS} \texttt{cublasCgetrfBatched} Cholesky decomposition, followed by \texttt{cublasCgetriBatched} that implements  forward and backward substitution operations to obtain $\mathbf{A}^{-1}$. 
Since the local channel matrix $\mathbf{H}_c$ is assumed to remain constant for $K$ OFDM symbols, 
we store $\mathbf{A}^{-1}$ for a given OFDM symbol $k$, and reuse this matrix for all $K$ OFDM symbols.  
%
To compute the whitened vector $\mathbf{z}=\frac{1}{\beta^{\text{WF}}}\mathbf{A}^{-1}\mathbf{s}$, we first multiply   the transmit vector $\mathbf{s}$ 
with the matrix $\mathbf{A}^{-1}$ using  the \texttt{cublasCgemmBatched} function for a total of $N_\text{sc}\times K$ subcarriers.
We then calculate the precoding factor $\beta^{\text{WF}}$. As shown in~\eqref{eq:beta}, $\beta^{\text{WF}}$ depends on $\tr(\mathbf{A}^{-1})$ and $\|\bA^{-1}\|_F^2$, 
which involve sum reduction operations across the diagonal entries or all entries of matrix $\mathbf{A}^{-1}$. 
To resolve such data dependencies efficiently, we design a customized kernel function to calculate~$\beta^{\text{WF}}$, where we  take advantage of fast local registers and shared memories for inter-thread communications. 
Specifically, we invoke this kernel with $N_\text{sc}$ thread-blocks to calculate $N_\text{sc}$ different $\beta^{\text{WF}}$ values  in parallel. 
In each thread-block, we generate $U\times U$ threads to access each entry of the $U\times U$ matrix~$\mathbf{A}^{-1}$, and perform inter-thread shuffle of local register values  within a \emph{warp} using \emph{warp shuffle}~\cite{warpshuffle}, and inter-thread communication across different \emph{warps} within this thread-block using shared memory, to realize the sum reductions required to compute~$\tr(\mathbf{A}^{-1})$ and $\|\bA^{-1}\|_F^2$. 
Analogously to the computations for $\mathbf{A}^{-1}$, we can reuse the parameter~$\beta^{\text{WF}}$ across $K$ OFDM symbols, and compute the whitened vector $\mathbf{z}$. 
For PD-WF, whitening happens at the master GPU in a centralized manner, and therefore we need to broadcast the whitened vector $\mathbf{z}$  to all  GPUs using NCCL \texttt{ncclBcast}. We finally compute the local precoding vector $\mathbf{x}_c=\mathbf{H}^H_c\mathbf{z}$ by \texttt{cublasCgemmBatched} function for all $N_\text{sc}\times K$ subcarriers at each GPU in a decentralized fashion.

\subsubsection{FD-WF Implementation}
%
For FD-WF, we use \emph{cuBLAS} and customized kernels as for PD-WF in order to implement the local WF precoder corresponding to $\mathbf{B}_c$ BS antennas with the power constraint $\rho_c^2=\frac{\rho^2}{C}$. To invoke the FD-WF precoder, we broadcast the transmit vectors $\mathbf{s}$ to the $C$ MPI processes, each running a local WF precoder on a separate GPU to compute the  local precoding vectors $\mathbf{x}_c$ in parallel.

\subsection{Implementation Results}
%
%
\fref{fig:berandtp} shows the latency and throughput results of PD-WF and FD-WF measured on the multi-GPU system for  various BS antenna configurations and $U=16$ UEs. 
%
For all configurations, we set $N_\text{sc}=1200$, 
$K=7$ corresponding to a slot-frame of 20~MHz LTE signal with OFDM and 64-QAM transmission.

In the top row of \fref{fig:berandtp} , we fix the number of BS antennas to $B=256$, and increase the number of clusters $C=2, 4, 8$ (and, hence, the number of used GPUs) while decreasing the cluster size $B_c=128, 64, 32$.  By decreasing $B_c$, the throughput of PD-WF and FD-WF precoding increases as less local workload is  processed in parallel. FD-WF achieves higher data rate than PD-WF, since FD-WF only requires to broadcast the transmit vector $\mathbf{s}$ which scales with $N_\text{sc}\times K \times U$, while PD-WF requires a higher amount of message passing, which includes (i) gathering of  local Gram matrices $\mathbf{G}_c$ (scaling with $N_\text{sc}\times U\times U$) and (ii) broadcasting of whitened vector $\mathbf{z}$ (scaling with $N_\text{sc}\times K \times U$).

In the bottom row of \fref{fig:berandtp}, we fix the number of antennas per cluster to $B_c=32$, and increase $B=64, 128, 256$ by scaling the number of clusters $C=2, 4, 8$. We observe that the throughput only degrades slightly with increasing $B$ and $C$ for both PD-WF and FD-WF, indicating that the message-passing latency remains nearly constant; this is a result of the direct GPU-to-GPU gathering/broadcasting communications realized by NCCL. 
This observation also implies that we can increase the number of BS antennas with only a small loss in throughput using the proposed decentralized feedforward architecture. 
For all configurations show in \fref{fig:berandtp}, our designs achieve throughputs in the Gb/s regime with latencies below $1$\,ms. We also see that FD-WF outperforms PD-WF in terms of throughput due to the reduced amount of message passing, while PD-WF achieves superior error-rate performance.

\section{Conclusions}
We have proposed two novel feedforward architectures and corresponding decentralized precoding algorithms based on the linear Wiener filter (WF) precoder. We have demonstrated that the partially-decentralized (PD) WF precoder  achieves the error-rate performance of the centralized WF precoder, while significantly reducing the interconnect and chip input/output bandwidths. To further reduce the interconnect bandwidth, we have proposed a fully-decentralized (FD) WF precoder that incurs only a small error-rate performance loss compared to the PD-WF precoder. 
To showcase the efficiency and scalability of PD-WF and FD-WF to large antenna arrays, we have presented implementations on a multi-GPU system. Our results demonstrate that throughputs in the Gb/s regime at latencies below $1$\,ms are achievable. These results indicate that the proposed decentralized precoding are a solution to combat the interconnect and complexity bottlenecks while being able to fully exploit the spectral efficiency and link reliability advantages provided by massive MU-MIMO systems.

There are many avenues for future work. 
A theoretical analysis of the optimal regularization parameter $\tau_c$ for FD-WF precoding in \fref{eq:regularization} is an open research question. 
Combining  decentralized feedforward precoding with data detection as in~\cite{isit17} may further reduce the processing latency and increase the throughput as a large number of quantities can be re-used between the uplink and downlink. 
The development and analysis of feedforward architectures for cell-free massive MU-MIMO as put forward in~\cite{ngo2017cell} is part of ongoing work.

\appendices

\section{Proof of \fref{thm:WFprecoder}}
\label{app:WFprecoderproof}
The precoder resulting from \fref{sec:WFproblem} is known as the Wiener filter (WF) precoder~\cite{wfprecoding} and can be derived as follows. 
Let us first form the Lagrangian
\begin{align*}
\setL(\bP,\beta,\lambda) =\, &  \Ex{\bms}{\|\bms-\beta\bH\bP\bms\|^2} + \beta^2 U \No \\
 & + \lambda(\Ex{\bms}{\|\bmx\|^2} - \rho^2).
\end{align*}
We can now formulate the optimality conditions for $\bP$ and $\beta$ by using the Wirtinger derivative as follows. For the precoding matrix $\bP$, we have the following optimality condition:
\begin{align}
\frac{\delta}{\delta \bP^H} \setL(\bP,\beta,\lambda)   = \bZero  \implies 
\beta^2 \bH^H \bH \bP + \lambda\bP  =  \beta \bH^H. \label{eq:optcondP}
\end{align}
For the precoding factor $\beta$, we compute $\frac{\delta}{\delta \beta^* } \setL(\setP,\beta,\lambda)  = 0$ and obtain  the following optimality condition:
\begin{align}
\beta \tr(\bP^H\bH^H\bH\bP) +\beta \frac{U \No}{\Es} & = \tr(\bH^H\bP^H). \label{eq:optcondbeta}
\end{align}
From the power constraint, it follows that 
\begin{align}
\Ex{\bms}{\|\bmx\|^2}   = \rho^2 & \implies
 \tr(\bP^H\bP) = \frac{\rho^2}{\Es}. \label{eq:optcondpower}
\end{align}
To derive the optimal value for the Lagrange multiplier $\lambda$, we apply the following steps the optimality condition in \fref{eq:optcondP}:
\begin{align}
\beta^2 \bH^H \bH \bP + \lambda\bP & =  \beta \bH^H \notag \\ 
\beta^2 \bH^H \bH \bP\bP^H + \lambda\bP\bP^H & =  \beta \bH^H\bP^H  \notag \\
\beta^2 \tr(\bH^H \bH \bP\bP^H) + \lambda\tr(\bP\bP^H) & =  \beta \tr(\bH^H\bP^H) \notag \\
\beta^2 \tr(\bP^H\bH^H \bH \bP) + \lambda \frac{\rho^2}{\Es} & =  \beta \tr(\bH^H\bP^H), \label{eq:optcondtrick}
\end{align}
where the last step results from \fref{eq:optcondpower}. We now multiply both sides of the optimality condition in \fref{eq:optcondbeta} with $\beta$ to obtain 
\begin{align}
\beta \tr(\bP^H\bH^H\bH\bP)\Es +\beta U \No & = \tr(\bH^H\bP^H)\Es \notag \\
\beta^2 \tr(\bP^H\bH^H\bH\bP) +\beta^2 \frac{U \No}{\Es} & = \beta \tr(\bH^H\bP^H). \label{eq:papapap1} 
\end{align}
Subtracting \fref{eq:papapap1}  from \fref{eq:optcondtrick} yields the Lagrange multiplier
\begin{align}
\lambda = \frac{U\No}{\rho^2}. \label{eq:lagrange}
\end{align}
From \fref{eq:optcondP} and \fref{eq:lagrange}, it follows that the WF precoding matrix is given by $\bP^\text{WF} = \frac{1}{\beta^\text{WF}}\bQ$ with the matrix
\begin{align*}
\bQ = \left(\bH^H\bH + \frac{U\No}{\rho^2}\bI \right)^{-1}\bH^H.
\end{align*}
The remaining piece is to identify the WF precoding factor $\beta^\text{WF}$. To this end, we plug $\bP^\text{WF}$ into~\fref{eq:optcondpower}, which leads to
\begin{align*}
\frac{1}{\beta^2} \tr(\bQ^H\bQ)  = \frac{\rho^2}{\Es} \implies
\frac{1}{\beta^\text{WF}}  = \frac{\rho}{\sqrt{\tr(\bQ^H\bQ) \Es}}.
\end{align*}

\section{Proof of \fref{lem:precodingfactor}}
\label{app:precodingfactor}
To reduce the computational complexity of computing $\beta^\text{WF}$ in \fref{eq:precodingfactor}, we first use the matrix inversion lemma~\cite{woodbury} to arrive at an equivalent expression of \fref{eq:originalQmatrix} given by
\begin{align*} 
\bQ^\text{WF} = \bH^H \left(\bH\bH^H + \kappa^\text{WF}\bI_U  \right)^{\!-1},
\end{align*}
which requires the inversion of an $U\times U$ matrix. By precomputing the $U\times U$ Gram matrix $\bG=\bH\bH^H$ and inverting the regularized Gram matrix defined as  $\bA = \bG + \kappa^\text{WF} \bI_U$, we have
\begin{align} \label{eq:WFprecoderinvertedcompact} 
\bQ^\text{WF} = \bH^H \bA^{-1}
\end{align}
and consequently, $\tr(\bQ^H\bQ) = \tr\left(  \bA^{-1} \bG \bA^{-1}  \right)$. A direct evaluation of this expression still requires two matrix-matrix multiplications of dimension  $U\times U$. We can further reduce complexity by noting that the following equivalence holds
\begin{align*}
\tr\left(  \bA^{-1} \bG \bA^{-1} \right) =  \tr\left(  \bA^{-1} \right) - \kappa^\text{WF}  \|\bA^{-1}\|_F^2,
\end{align*}
where we used a Searle-type identity~\cite{PP2012} and a matrix version of partial fraction expansion to finally obtain \fref{eq:beta}.

\balance
\bibliographystyle{IEEEbib}
\bibliography{MIMO}

\end{document}